\documentclass[12pt]{article}
\usepackage{amssymb}
\renewcommand{\theequation}{\arabic{section}.\arabic{equation}}

\begin{document}



\def\a{\alpha}
\def\b{\beta}
\def\d{\delta}
\def\e{\epsilon}
\def\g{\gamma}
\def\h{\mathfrak{h}}
\def\k{\kappa}
\def\l{\lambda}
\def\o{\omega}
\def\p{\wp}
\def\r{\rho}
\def\t{\theta}
\def\s{\sigma}
\def\z{\zeta}
\def\x{\xi}
 \def\A{{\cal{A}}}
 \def\B{{\cal{B}}}
 \def\C{{\cal{C}}}
\def\D{\Delta}
\def\G{\Gamma}
\def\K{{\cal{K}}}
\def\O{\Omega}
\def\P{{\cal{P}}}
\def\L{\Lambda}
\def\f{E_{\tau,\eta}(sl_2)}
\def\E{E_{\tau,\eta}(sl_n)}
\def\Zb{\mathbb{Z}}
\def\Cb{\mathbb{C}}

\def\R{\overline{R}}

\def\beq{\begin{equation}}
\def\eeq{\end{equation}}
\def\bea{\begin{eqnarray}}
\def\eea{\end{eqnarray}}
\def\ba{\begin{array}}
\def\ea{\end{array}}
\def\no{\nonumber}
\def\le{\langle}
\def\re{\rangle}
\def\lt{\left}
\def\rt{\right}

\newtheorem{Theorem}{Theorem}
\newtheorem{Definition}{Definition}
\newtheorem{Proposition}{Proposition}
\newtheorem{Lemma}{Lemma}
\newtheorem{Corollary}{Corollary}
\newcommand{\proof}[1]{{\bf Proof. }
        #1\begin{flushright}$\Box$\end{flushright}}

\baselineskip=20pt

\newfont{\elevenmib}{cmmib10 scaled\magstep1}
\newcommand{\preprint}{
   \begin{flushleft}
     \elevenmib Yukawa\, Institute\, Kyoto\\
   \end{flushleft}\vspace{-1.3cm}
   \begin{flushright}\normalsize  \sf
     YITP-04-21\\
     {\tt hep-th/0404089} \\ April 2004
   \end{flushright}}
\newcommand{\Title}[1]{{\baselineskip=26pt
   \begin{center} \Large \bf #1 \\ \ \\ \end{center}}}
\newcommand{\Author}{\begin{center}
   \large \bf Wen-Li Yang${}^{a,b}$
    ~ Alexander Belavin${}^c$ and~Ryu Sasaki${}^d$\end{center}}
\newcommand{\Address}{\begin{center}

     ${}^a$ Institute of Modern Physics, Northwest University
     Xian 710069, P.\,R. China\\
     ~~\\
     ${}^b$ Department of Mathematics, University of Queensland,
     Brisbane, QLD 4072, Australia\\
     ~~\\
     ${}^c$ Landau Institute for Theoretical Physics, Chernogolovka,
     142432, Russia\\
     ~~\\
     ${}^d$ Yukawa Institute for Theoretical Physics,\\
     Kyoto University, Kyoto 606-8502, Japan
\end{center}}
\newcommand{\Accepted}[1]{\begin{center}
   {\large \sf #1}\\ \vspace{1mm}{\small \sf Accepted for Publication}
   \end{center}}

\preprint
\thispagestyle{empty}
\bigskip\bigskip\bigskip

\Title{Central elements of the elliptic $\Zb_n$ monodromy matrix
algebra at roots of unity} \Author

\Address
\vspace{1cm}

\begin{abstract}
The central elements of the algebra of monodromy matrices
associated  with the $\Zb_n$ R-matrix are studied. When the
crossing parameter $w$ takes a special rational value
$w=\frac{n}{N}$, where $N$ and $n$ are positive coprime integers,
the center is substantially larger than that in the generic case
for which the ``quantum determinant" provides the center.  In the
trigonometric limit, the situation corresponds to the quantum
group at roots of unity. This is a higher rank generalization of
the recent results by Belavin and Jimbo.

\vspace{1truecm}

\end{abstract}
\newpage
\section{Introduction}
\label{intro} \setcounter{equation}{0} The algebra of the
monodromy matrix (or Yang-Baxter algebra) is generated by
noncommutative matrix elements of the L-operator $L(u)$ satisfying
a quadratic ``RLL" relation \bea
R_{12}(u-v)L_1(u)L_2(v)=L_2(v)L_1(u)R_{12}(u-v),\label{RLL}\eea
where $L_1(u)=L(u)\otimes 1$, $L_2(u)=1\otimes L(u)$ and  $R(u)$
is the R-matrix, a solution of quantum Yang-Baxter equation \bea
R_{12}(u-v)R_{13}(u)R_{23}(v)= R_{23}(v)R_{13}(u)R_{12}(u-v).
\label{YBE-V}\eea We adopt the standard notation: $R_{ij}(u)$ is
an embedding operator of R-matrix in the tensor space
$\Cb^n\otimes\Cb^n\cdots$, which acts as identity on the factor
spaces except for the $i$-th and $j$-th ones. If $R(u)$ is a
rational R-matrix, the algebra after a proper specialization is a
Yangian  algebra \cite{Dri85}. For a trigonometric R-matrix, the
algebra is closely connected with a quantum group \cite{Dri86,
Jim85}. If $R(u)$ is an elliptic R-matrix, the algebra gives rise
to a Sklyanin algebra \cite{Skl83,Hou89}.

In this paper, we will address the issue of  the center of an
elliptic algebra associated with the $\Zb_n$ R-matrix \cite{Bel81}
when the crossing parameter takes rational values. In the
trigonometric limit, it corresponds to a quantum group at roots of
unity. So here we call this case  the ``{\it roots of unity \/}"
case as its trigonometric limit. As is well known, for the
trigonometric algebra (as a degenerate algebra of the elliptic
algebra) when the crossing parameter (deformation parameter) is in
roots of unity case the center is much larger than that of the
generic case \cite{Con90, Tar93}. Recently for the simplest case
of the $\Zb_2$ R-matrix (or eight-vertex R-matrix) \cite{Bax86},
the structure of the center in the ``roots of unity" case was
clarified  \cite{Bel02}. Here, we will consider the question with
a generic case $n\geq 2$.

The paper is organized as follows. In section 2 we give the
definition of the elliptic algebra and introduce appropriate
notation. We formulate the fundamental Boltzmann weights of the
$A_{n-1}$ type interaction-around-a-face (IRF) model (or the SOS
model) and give the corresponding face-vertex correspondence
relation in section 3. In section 4, we construct the symmetric
fusion procedure for the R-matrices, Boltzmann weights and
intertwining vectors, which are useful to handle  the center of
the algebra. Finally, we give the center of the elliptic algebra
with the  $\Zb_n$ R-matrix when the crossing parameter is in the
``roots of unity" case in section 5. Section 6 is for conclusions.

\section{The Algebra of Monodromy Matrices   }
 \label{AMM} \setcounter{equation}{0}


Let us fix an integer $n$ ($n\geqslant 2$) and a complex number
$\tau$ such that $Im(\tau)>0$. Introduce the following elliptic
functions \bea &&\t\lt[
\begin{array}{c}
a\\b
\end{array}\rt](u,\tau)=\sum_{m=-\infty}^{\infty}
\exp\lt\{\sqrt{-1}\pi\lt[(m+a)^2\tau+2(m+a)(u+b)\rt]\rt\},\\
&&\t^{(j)}(u)=
\t\lt[\begin{array}{c}\frac{1}{2}-\frac{j}{n}\\[2pt]\frac{1}{2}
\end{array}\rt](u,n\tau),~~~
\s(u)=\t\lt[\begin{array}{c}\frac{1}{2}\\[2pt]\frac{1}{2}
\end{array}\rt](u,\tau),\label{Function}\\
&&\bar{\t}^{(j)}(u)=\exp\lt\{\sqrt{-1}\pi
u\rt\}\,\t^{(j)}(u).\label{Function1}
 \eea Among them the $\s$-function\footnote{Our
$\s$-function is the $\vartheta$-function $\vartheta_1(u)$
\cite{Whi50}. It has the following relation with the {\it
Weierstrassian\/} $\s$-function if denoted it by $\s_w(u)$:
$\s_w(u)\propto e^{\eta_1u^2}\s(u)$,
$\eta_1=\pi^2(\frac{1}{6}-4\sum_{n=1}^{\infty}\frac{nq^{2n}}{1-q^{2n}})
$ and $q=e^{\sqrt{-1}\tau}$.} satisfies the following
identity:\bea
&&\s(u+x)\s(u-x)\s(v+y)\s(v-y)-\s(u+y)\s(u-y)\s(v+x)\s(v-x)\no\\
&&~~~~~~=\s(u+v)\s(u-v)\s(x+y)\s(x-y).\no\eea

Let $\lt\{e_{i}~|~i=1,2,\cdots,n\rt\}$  be the orthonormal basis
of the vector space $\Cb^n$ such that $\langle e_i,~e_j
\rangle=\d_{ij}$. Let $R(u)\in End(\Cb^n\otimes\Cb^n)$ be the
$\Zb_n$ R-matrix given by  \bea
R(u)=\sum_{i,j,k,l}R^{kl}_{ij}(u)E_{ik}\otimes
E_{lj},\label{Belavin-R}\eea in which $E_{ij}$ is the matrix with
elements $(E_{ij})^l_k=\d_{jk}\d_{il}$. The coefficient functions
are \cite{Ric86, Jim87} \bea R^{kl}_{ij}(u)=\lt\{
\begin{array}{ll}
\frac{h(u)\t^{(i-j)}(u+w)}{\t^{(i-k)}(w)\t^{(k-j)}(u)}&{\rm
if}~i+j=k+l~{\rm mod}~n,\\[6pt]
0&{\rm otherwise,}
\end{array}\rt.\label{Belavin-R1}
\eea where a complex parameter $w$ is called the {\it crossing
parameter\/}. We have set \bea
h(u)=\frac{\prod_{j=0}^{n-1}\t^{(j)}(u)}
{\prod_{j=1}^{n-1}\t^{(j)}(0)}.\no\eea The R-matrix satisfies the
quantum Yang-Baxter equation (\ref{YBE-V}),  unitarity and
crossing-unitarity relations \cite{Ric86}.  It should be remarked
that our R-matrix given in (\ref{Belavin-R}) and
(\ref{Belavin-R1}) is the same as that of Ref.\cite{Jim87}, and
that our R-matrix $R(u)$ is related to the R-matrix $S(u)$ of
Richey and Tracy \cite{Ric86} by $R_{12}(u)=\frac{e^{\sqrt{-1}\pi
u}}{n}S^{t_1t_2}_{12}(u)$, where $t_i$ denotes the transposition
in the $i$-th space. Consequently,  the R-matrix enjoys the
following  property:
\begin{Lemma}
The R-matrix given in (\ref{Belavin-R}) and (\ref{Belavin-R1})
satisfies \bea R(w)=M\times P_2^{(+)},\label{Fusion} \eea where
$M$ is a non-degenerate matrix and  $P_2^{(+)}$ is the two-body
symmetrization operator given by $P_2^{(+)}=\frac{1}{2}(1+P_{12})$
in terms of the permutation operator $P_{12}$: $P_{12}(e_i\otimes
e_j)=e_j\otimes e_i$.
\end{Lemma}
This lemma tells that \bea Ker\lt(R(w)\rt)={\rm
the~antisymmetric~subspace~of
}~\Cb^n\otimes\Cb^n.\label{Fusion-P}\eea The above property
enables us to construct symmetric fusion of the R-matrix in higher
tensor space \cite{Che85}  and the intertwining vector in section
4.

\begin{Definition}
The algebra of monodromy matrices $\cal{A}$ is an associative
algebra generated  by the matrix elements of the $L$-operator
$L_{ij}(u)$, $i,j=1,\cdots,n$ with the defining relations
(\ref{RLL}).
\end{Definition}

One can introduce a natural coproduct $\D$: $\cal{A}\rightarrow
\cal{A}\otimes\cal{A}$ and counit $\e$: $\cal{A}\rightarrow \Cb$
\bea \D(L_{ij}(u))=\sum_{k}L_{ik}(u)\otimes L_{kj}(u),~~
\e(L_{ij})=\d_{ij},\no \eea making $\cal{A}$ a bi-algebra. Certain
realization of the algebra $\A$ can be expressed in terms of the
$\Zb_n$ Sklyanin algebra \cite{Skl83,Hou89} which is an elliptical
generalization of quantum group $U_q(gl_n)$.

In this paper, we investigate the center of the algebra $\A$,
i.e., the set of the elements commuting with all the elements of
$\A$. For a generic value of the crossing parameter $w$, the
so-called quantum determinant of the L-operator ($det_qL(u)$)
generates the center of the algebra $\A$ \cite{Hou90}. However, it
is known \cite{Bel02} that $\A$ for the eight-vertex model case
($n=2$) the situation is quite different when $w=\frac{2}{N}$, $N$
is a positive odd integer. In this case, in addition to the
``quantum determinant" $det_qL(u)$, there exist much more central
elements in $\A$. For the generic case of $n\geqslant 2$, the same
phenomenon has been observed for $\A$ in the trigonometric limit
\cite{Con90, Tar93}. We shall study the {\it extra} central
elements of $\A$ with a generic $n\geqslant 2$ in the ``roots of
unity" case: $w=\frac{n}{N}$, where $N$ and $n$ are  positive
coprime integers. Hereafter, we restrict the crossing parameter to
the ``roots of unity" case $w=\frac{n}{N}$, unless otherwise
stated.

\section{$A^{(1)}_{n-1}$ Face Model and Face-Vertex Correspondence}
\label{FVC} \setcounter{equation}{0} In this section, we formulate
the fundamental Boltzmann weights of the $A_{n-1}$ type SOS face
model and give the face-vertex correspondence relation.

An ordered pair $(a,b)$ $(a,b\in \Cb^n)$ is called {\it
admissible\/} if \bea
b-a=e_i,~~i=1,\cdots,n,\label{Admissible}\eea and denoted by
$a\stackrel{1}{\rightarrow}b$. A path
$p=(a^{(0)},a^{(1)},\cdots,a^{(l)})\in (\Cb^n)^{l+1}$ is called an
$l$-path from $a$ to $b$ if $a^{(0)}=a$, $a^{(l)}=b$, and all
pairs $(a^{(i-1)},a^{(i)})$ for $i=1,\cdots,l$ are admissible.

\begin{Definition}
An ordered pair $(a,b)$ $(a,b\in \Cb^n)$ is called {\it
l-admissible\/} if there exists an l-path from $a$ to $b$, and
denoted by $a\stackrel{l}{\rightarrow}b$.
\end{Definition}

Let $W\lt(\lt.\begin{array}{ll}a&b\\c&d\end{array}\rt|\,u\rt)$
denote the Boltzmann weight corresponding to a configuration
$a,\,b,\,d$, $c~\in \Cb^n$ around a face, ordered clockwise from
the NW corner. We call that
$W\lt(\lt.\begin{array}{ll}a&b\\c&d\end{array} \rt|\,u\rt)$ is
{\it admissible\/} if $(a,b)$, $(b,d)$, $(a,c)$, $(c,d)$ are all
{\it admissible\/}. By definition non-admissible weights are set
to $0$. The non-vanishing weights are given as follows \bea
&&W\lt(\lt.\begin{array}{ll}a&a+e_{\mu}\\a+e_{\mu}&a+2e_{\mu}\end{array}
\rt| u\rt)=\frac{\s(u+w)}{\s(w)},\label{W-matrix1}\\
[5pt]
&&W\lt(\lt.\begin{array}{ll}a&a+e_{\mu}\\a+e_{\mu}&a+e_{\mu}+e_{\nu}
\end{array}\rt| u\rt)=\frac{\s(a_{\mu\nu}w -u)}{\s(a_{\mu\nu}w)},~~\mu\neq \nu,
\label{W-matrix2}\\[5pt]
&&W\lt(\lt.\begin{array}{ll}a&a+e_{\nu}\\a+e_{\mu}&a+e_{\mu}+e_{\nu}
\end{array}\rt| u\rt)=\frac{\s(u)\s(a_{\mu\nu}w+w)}{\s(w)\s(a_{\mu\nu}w)},~~\mu\neq
\nu,\label{W-matrix3}\\[5pt]
&&a_{\mu\nu}=\langle
a,e_{\mu}-e_{\nu}\rangle,~~\mu,\nu=1,\cdots,n.\eea The weights
given in (\ref{W-matrix1})-(\ref{W-matrix3}) satisfy the
star-triangle relation (or dynamical Yang-Baxter equation)
\cite{Jim87} \bea
&&\sum_{g}\,W\lt(\lt.\begin{array}{ll}a&b\\f&g\end{array}
\rt|\,u+v\rt)\, W\lt(\lt.\begin{array}{ll}f&g\\e&d\end{array}
\rt|\,u\rt)\,
W\lt(\lt.\begin{array}{ll}b&c\\g&d\end{array} \rt|\,v\rt)\no\\
&&~~~~=\sum_{g}\,W\lt(\lt.\begin{array}{ll}a&g\\f&e\end{array}
\rt|\,v\rt)\, W\lt(\lt.\begin{array}{ll}a&b\\g&c\end{array}
\rt|\,u\rt)\, W\lt(\lt.\begin{array}{ll}g&c\\e&d\end{array}
\rt|\,u+v\rt).\label{STR} \eea

For a generic $\l\in \Cb^n$, define \bea
&&\l_i=\langle\l,e_i\rangle,
~~\bar{\l}_i=\langle\l,\bar{e}_i\rangle,~~\bar{e}_i=e_i-\frac{1}{n}\sum_ke_k,
\label{Def1}\\
&&\l_{ij}=\l_i-\l_j=\bar{\l}_i-\bar{\l}_j,~~i,j=1,\cdots,n.
\label{Def2}\eea Let us introduce an intertwining vector---an
$n$-component column vector $\phi_{\l,\l+e_{j}}(u)$ whose $k$-th
element is \bea
\phi^{(k)}_{\l,\l+e_{j}}(u)=\bar{\t}^{(k)}(u-nw\bar{\l}_j).\label{Interwiner}\eea
One can easily verify the following properties of the intertwining
vector from the definition (\ref{Interwiner}) and equation
(\ref{Def1}).
\begin{Lemma} \label{Shift}
The intertwining vector defined by (\ref{Interwiner}) satisfies
the following relation \bea
&&\phi_{\l+e_j,\l+e_i+e_{j}}(u)=\phi_{\l,\l+e_{i}}(u+w-n\d_{ij}w),~~i,j=1,\cdots,n,\label{F-S-P}\\
&&\phi_{\l,\l+e_i}(u+n)=\phi_{\l,\l+e_i}(u).\label{G-P}\eea
\end{Lemma}
The intertwining vector (\ref{Interwiner}) is different  from that
of Jimbo et al \cite{Jim87} by the definition of the
$\bar{\t}$-function given  in (\ref{Function1}). The present
definition is preferred because of its  periodicity (\ref{G-P})
which will play an important role in section 5. From the
definition of the Boltzmann weights
(\ref{W-matrix1})-(\ref{W-matrix3}) and using the face-vertex
correspondence  obtained in \cite{Jim87}, we have the following
relation.
\begin{Theorem} \label{VF} The intertwining vector satisfies
\bea &&R_{12}(u-v) \phi_{\l,\l+e_{i}}(u)\otimes
\phi_{\l+e_{i},\l+e_{i}+e_{j}}(v)\no\\
&&~~~~=\sum_{k}\,W\lt.\lt(\begin{array}{ll}\l&\l+e_k\\\l+e_i&\l+e_i+e_j\end{array}
\rt|\,u-v\rt) \phi_{\l+e_{k},\l+e_{i}+e_{j}}(u)\otimes
\phi_{\l,\l+e_{k}}(v). \label{Face-vertex}\eea
\end{Theorem}

Then the Yang-Baxter equation of the $\Zb_n$ R-matrix $R(u)$
(\ref{YBE-V}) is equivalent to the star-triangle relation
(\ref{STR}).

\section{Fusion Procedure}
 \label{FP} \setcounter{equation}{0}
The fusion procedure of the R-matrix and the intertwining vector
is essential  in handling the central elements of the algebra
$\A$.
\subsection{The generic $w$ case}
In this subsection, we consider the case that the crossing
parameter $w$ takes a generic value. Let us introduce operators
${\cal{L}}_{1\cdots N}(u)$ and ${\cal{R}}_{1\cdots N}(u)$  by \bea
&&{\cal{R}}_{1\cdots N}(u)\equiv {\cal{R}}_{1\cdots
N;0}(u)=R_{10}(u+(N-1)w)\cdots
R_{N0}(u),\label{F-R}\\
&&{\cal{L}}_{1\cdots N}(u)=L_1(u+(N-1)w)\cdots L_N(u).\label{F-L}
\eea We consider ${\cal{L}}_{1\cdots N}(u)$ ( ${\cal{R}}_{1\cdots
N}(u)$) as an operator acting on the so-called ``{\it auxiliary
space\/}" $V_1\otimes\cdots\otimes V_N$, $V_i=\Cb^n$ with the
entries belonging to $\A$ (with the entries of the operator
${\cal{R}}_{1\cdots N}(u)$ acting on an additional $n$-dimensional
space denoted by $0$-th tensor space). Define an operator $S_N$
acting on $V_1\otimes\cdots\otimes V_N$ as follow \bea
S_N=\prod_{i=1}^{N-1}\prod_{j=i+1}^{N}\,R_{ij}\lt((j-i)w\rt),
\label{S-op}\eea where both indices grow from left to right. For
example, if $N=3$, then \bea
S_3=R_{12}(w)R_{13}(2w)R_{23}(w).\no\eea
\begin{Proposition} \label{Invar}
$Ker(S_N)$ is an invariant subspace in $V_1\otimes\cdots\otimes
V_N$ of the operators ${\cal{L}}_{1\cdots N}(u)$ and
${\cal{R}}_{1\cdots N}(u)$.
\end{Proposition}
{\it Proof\/}. Using the Yang-Baxter equation (\ref{YBE-V}) and
$RLL$ relation (\ref{RLL}), one can derive \bea
&&S_N{\cal{R}}_{1\cdots N}(u)={\cal{R}}_{N\cdots 1}(u)S_N,\no\\
&&S_N{\cal{L}}_{1\cdots N}(u)={\cal{L}}_{N\cdots 1}(u)S_N,\no\eea
where ${\cal{R}}_{N\cdots 1}(u)$ and ${\cal{L}}_{N\cdots 1}(u)$
differ from ${\cal{R}}_{1\cdots N}(u)$ and ${\cal{L}}_{1\cdots
N}(u)$ by permutation of all factors to the opposite order. Hence,
${\cal{R}}_{1\cdots N}(u)\,Ker(S_N)\subset Ker(S_N)$,
${\cal{L}}_{1\cdots N}(u)\,Ker(S_N)\subset
Ker(S_N)$.~~~~~~~$\square$

Now, let us consider the subspace $Ker(S_N)$. Define \bea
W_0=\Sigma^{N-1}_{i=1}~ Ker\lt(R_{i,i+1}(w)\rt)\,\subset
V_1\otimes\cdots\otimes V_N.\label{KC}\eea Using the Yang-Baxter equation
(\ref{YBE-V}), one can show that the factors of $S_N$ can be
permuted into the order such that $R_{i,i+1}(w)$ for a given
$i=1,\cdots,N-1$ is in the rightmost position. For example, if
$N=3$ and $i=1$, \bea
S_3=R_{12}(w)R_{13}(2w)R_{23}(w)=R_{23}(w)R_{13}(2w)R_{12}(w).\no\eea
So, $Ker(S_N)\supset W_0$. Comparing the dimensions of the
subspaces in the trigonometric limit, one can derive that
$Ker(S_N)=W_0$ for the generic crossing parameter $w$. However,
when the crossing parameter $w$ is in the ``roots of unity" case
$w=\frac{n}{N}$, the subspace $Ker(S_N)$ will be larger than
$W_0$.

It follows from Proposition \ref{Invar} that $W_0$ is invariant
under  ${\cal{L}}_{1\cdots N}(u)$ and ${\cal{R}}_{1\cdots N}(u)$.
Thus these operators can be defined in the quotient space
$W_0^{\perp}$ which is defined as follow \bea
W_0^{\perp}=V_1\otimes\cdots\otimes V_N/W_0.\label{KC1}\eea This
also enables one to
construct the symmetric fusion of the R-matrix and the L-operator
in higher tensor space \cite{Che85}. Moreover, the property
(\ref{Fusion-P}) and the definitions of $W_0$ (\ref{KC}) and
$W_0^{\perp}$ (\ref{KC1})
imply that \bea
&&W_0^{\perp}= {\rm
the~symmetric~subspace~of~}
V_1\otimes\cdots\otimes V_N,\label{SYM}\\
&&{\rm dim}(W_0^{\perp})=\frac{(N+n-1)!}{N!(n-1)!},~~ {\rm
dim}(W_0)=n^N-\frac{(N+n-1)!}{N!(n-1)!}. \eea Define a complete
symmetrizer $P_N^{(+)}$ which is given in terms of the element of
$N$-body permutation group denoted by $\P_N$ as follows: \bea
P_N^{(+)}=\frac{1}{N!}\lt\{\sum_{P\in\P_N}\,P
\rt\}.\label{SYM1}\eea Proposition \ref{Invar} and the equation
(\ref{SYM}) lead to the following fusion properties.
\begin{Lemma}
\bea &&P^{(+)}_N{\cal{R}}_{1\cdots N;0}(u)
=P^{(+)}_N{\cal{R}}_{1\cdots
N;0}(u)P^{(+)}_N,\label{F-PR}\\
&&P^{(+)}_N{\cal{L}}_{1\cdots N}(u)=P^{(+)}_N{\cal{L}}_{1\cdots
N}(u)P^{(+)}_N.\label{F-PL}\eea
\end{Lemma}

Hence, one can define the {\it symmetric\/} fused $R$-matrix
${\cal{R}}^{(s)}_{1\cdots N}$ and $L$-operator
${\cal{L}}^{(s)}_{1\cdots N}$ acting in $W_0^{\perp}$ as follows
\bea &&{\cal{R}}^{(s)}_{1\cdots N}(u)\equiv
{\cal{R}}^{(s)}_{1\cdots N;0}(u) =P^{(+)}_N{\cal{R}}_{1\cdots
N;0}(u)P^{(+)}_N,\label{F-R1}\\
&&{\cal{L}}^{(s)}_{1\cdots N}(u)=P^{(+)}_N{\cal{L}}_{1\cdots
N}(u)P^{(+)}_N.\label{F-L1}\eea

For an {\it N-admissible\/} pair $(a,b)$, we choose one of its
$N$-paths from $a$ to $b$: $p=
(a,a+e_{i_1},a+e_{i_1}+e_{i_2},\cdots,a+\sum_{l=1}^Ne_{i_l})$,
$b=a+\sum_{l=1}^Ne_{i_l}$. Let us introduce a fused intertwining
vector $\Phi_{p;a,b}(u)$ as follows \bea &&
\phi^{i_1,\cdots,i_N}_{N;a,b}(u)=\phi_{a,a+e_{i_1}}(u+(N-1)w)
\otimes\cdots\otimes\phi_{a+\sum_{l=1}^{N-1}e_{i_l},a+\sum_{l=1}^{N}e_{i_l}}(u),
\label{Ph-1}\\
&&\Phi_{p;a,b}(u)=P^{(+)}_{N}\lt\{
\phi^{i_1,\cdots,i_N}_{N;a,b}(u)\rt\}.\label{F-In} \eea Noting
Lemma \ref{Shift}, we obtain:
\begin{Proposition} \label{Prop-P}The fused intertwining vector given in
(\ref{F-In}) satisfies\bea \Phi_{p;a,b}(u)=\frac{1}{N!}
\lt\{P^{(+)}_N\lt\{\sum_{P\in \P_N
}\phi^{i_{p_1},\cdots,i_{p_N}}_{N;a,b}(u)\rt\}\rt\},
\label{Main-E}\eea where $P$ is an element of $\P_N$:\bea
P=\lt(\begin{array}{llll}1&2&\cdots&N\\p_1&p_2&\cdots&p_N\end{array}\rt).
\eea
\end{Proposition}
The proof of this Proposition is relegated  to Appendix.

\noindent Then, we have
\begin{Corollary} \label{ID}
The fused intertwining vector given in (\ref{F-In}) is independent
of the choice of the $N$-path from $a$ to $b$ for any {\it
$N$-admissible\/} pair ($a$, $b$).
\end{Corollary}
Because of the path independence, we hereafter  denote the fused
intertwining vector given in (\ref{F-In}) by $\Phi_{N;a,b}(u)$ for
any {\it $N$-admissible\/} pair ($a$, $b$) instead of
$\Phi_{p;a,b}(u)$. From the construction of the fused intertwining
vector, one easily has \bea {\rm Span}\lt\{\Phi_{N;a,b}(u)|~{\rm
for~all~} N-admissible~{\rm pairs} ~(a,b) \rt\}=W_0^{\perp}.\eea

Now we establish an equivalence between the symmetric fused
R-matrix (\ref{F-R1}) and Boltzmann weights considered in
\cite{Jim88} through the fused intertwining vector.
\begin{Proposition} \label{F-W-P}The fused intertwining vector satisfies the
following fused face-vertex correspondence relation \bea &&
{\cal{R}}^{(s)}_{1\cdots N;0}(u-v) \Phi_{N;a,b}(u)\otimes
\phi_{b,b+e_{j}}(v)\no\\
&&~~~~=\sum_{k}\,W_{N1}\lt(\lt.\begin{array}{ll}a&a+e_k\\b&b+e_j\end{array}
\rt|\,u-v\rt) \Phi_{N;a+e_{k},b+e_{j}}(u)\otimes
\phi_{a,a+e_{k}}(v). \label{Face-vertex1}\eea Here the fused
Boltzmann weights $
W_{N1}\lt(\lt.\begin{array}{ll}a&b\\d&c\end{array}\rt|u\rt)$ are
non-vanishing  only for a configuration $a,b,c,d\in \Cb^n$ such
that $a\stackrel{1}{\rightarrow}b$, $b\stackrel{N}{\rightarrow}c$,
$a\stackrel{N}{\rightarrow}d$, $d\stackrel{1}{\rightarrow}c$ and
the non-vanishing weights are given as follows\bea
W_{N1}\lt(\lt.\begin{array}{ll}a&b\\d&c\end{array}\rt|u\rt)=
\lt(\prod_{j=1}^{N-1}\frac{\s(u+jw)}{\s(w)} \rt)
\frac{\s(u+c_{\mu}w-b_{\nu}w)\prod_{\r\neq
\mu}\s(c_{\r}w-b_{\nu}w+w)}
{\prod_{\r}\s(a_{\r\nu}w+\d_{\r\nu}w)},\label{E-F} \eea when
$b-a=e_{\nu}$ and $c-d=e_{\mu}$.
\end{Proposition}
{\it Proof\/}. Suppose $(a,a^{(1)},\cdots,a^{(N-1)},b)$ is an
$N$-path from $a$ to $b$. By the fundamental face-vertex
correspondence relation of Theorem \ref{VF}, one can derive \bea
&&{\cal R}_{1\cdots N;0}
(u-v)\lt(\phi_{a,a^{(1)}}(u+(N-1)w)\otimes \cdots\otimes
\phi_{a^{(N-1)},b}(u)\rt)\otimes \phi_{b,b+e_{j}}(v)\no\\[5pt]
&&~~=\hspace{-5mm}\sum_{a'^{(1)},\cdots,a'^{(N-1)},b'}\hspace{-6mm}
W\hspace{-1.5mm}\lt(\lt.\begin{array}{ll}a&a'^{(1)}\\
a^{(1)}&a'^{(2)}\end{array}\rt|u-v+(N-1)w\hspace{-1mm}\rt)\hspace{-1.5mm}
W\hspace{-1.5mm}\lt(\lt.\begin{array}{ll}a^{(1)}&a'^{(2)}\\
a^{(2)}&a'^{(3)}\end{array}\rt|u-v+(N-2)w\hspace{-1mm}\rt)\no\\[5pt]
&&~~~~~~\times \cdots W\lt(\lt.\begin{array}{ll}a^{(N-2)}&a'^{(N-1)}\\
a^{(N-1)}&b'\end{array}\rt|u-v+w\rt)
W\lt(\lt.\begin{array}{ll}a^{(N-1)}&b'\\
b&b+e_{j}\end{array}\rt|u-v\rt)\no\\[5pt]
&&~~~~~~\times
\lt(\phi_{a'^{(1)},a'^{(2)}}(u+(N-1)w)\otimes\cdots\otimes
\phi_{b',b+e_{j}}(u)\rt) \otimes \phi_{a,a'^{(1)}}(v).\label{Eq}
\eea The definition of the fused intertwining vector (\ref{F-In})
leads to \bea \phi_{a,a^{(1)}}(u+(N-1)w)\otimes \cdots\otimes
\phi_{a^{(N-1)},b}(u)=\Phi_{N;a,b}(u)~~ {\rm mod}~
W_0.\label{Pro}\eea Multiplying equation (\ref{Eq}) by
$P^{(+)}_{N}$ from the left  and using Proposition \ref{Invar}, we
have \bea &&{\rm LHS}=P^{(+)}_{N}{\cal R}_{1\cdots N;0}(u-v)
(\Phi_{N;a,b}(u)+W_0)\otimes \phi_{b,\,b+e_{j}}(v)\no\\[3pt]
&&~~~={\cal R}^{(s)}_{1\cdots N;0}(u-v) \Phi_{N;a,b}(u)\otimes
\phi_{b,\,b+e_{j}}(v),\\[3pt]
&&{\rm RHS}=\sum_{a'^{(1)}}\hspace{-1mm}\lt\{\hspace{-1mm}
\sum_{a'^{(2)},\cdots,a'^{(N-1)},b'}\hspace{-7mm}
W\hspace{-1mm}\lt(\lt.\begin{array}{ll}\hspace{-2mm}a&\hspace{-2mm}a'^{(1)}\\
\hspace{-2mm}a^{(1)}&\hspace{-2mm}a'^{(2)}\end{array}\hspace{-1mm}\rt|\hspace{-1mm}
u\hspace{-1mm}-\hspace{-1mm}v\hspace{-1mm}+\hspace{-1mm}
(N\hspace{-1mm}-\hspace{-1mm}1)w\hspace{-1mm}\rt)\hspace{-1mm}
\cdots
\hspace{-1mm}W\hspace{-1mm}\lt(\lt.\begin{array}{ll}\hspace{-2mm}a^{(N-1)}&
\hspace{-3mm}b'\\\hspace{-2mm}b&
\hspace{-3mm}b+e_{j}\end{array}\hspace{-1mm}\rt|\hspace{-1mm}u\hspace{-1mm}
-\hspace{-1mm}v\hspace{-1mm}\rt)\rt.\no\\[3pt]
&&~~~~\times
\lt.P^{(+)}_{N}\lt\{\phi_{a'^{(1)},a'^{(2)}}(u+(N-1)w)\otimes\cdots\otimes
\phi_{b',b+e_{j}}(u)\rt\} \mbox{\vphantom{\Huge L}}\rt\}
\otimes \phi_{a,a'^{(1)}}(v)\no\\[3pt]
&&~~= \sum_{a'^{(1)}}\lt\{
\sum_{a'^{(2)},\cdots,a'^{(N-1)},b'}W\lt(\lt.\begin{array}{ll}a&a'^{(1)}\\
a^{(1)}&a'^{(2)}\end{array}\rt|u-v+(N-1)w\rt)\cdots\rt.\no\\[5pt]
&&~~~~\times\lt.W\lt(\lt.\begin{array}{ll}a^{(N-1)}&b'\\
b&b+e_{j}\end{array}\rt|u-v\rt)\rt\}
\Phi_{N;a'^{(1)},b+e_{j}}(u)\otimes \phi_{a,a'^{(1)}}(v)\no\\[5pt]
&&~~=\sum_{a'^{(1)}}
W_{N1}\lt(\lt.\begin{array}{ll}a&a'^{(1)}\\
b&b+e_{j}\end{array}\rt|u-v\rt)
\Phi_{N;a'^{(1)},b+e_{j}}(u)\otimes \phi_{a,a'^{(1)}}(v). \eea We
have used Corollary \ref{ID} in the second last equality and Lemma
2.2 of Ref.\cite{Jim88} in the last equality. The expression of
the fused Boltzmann weight
$W_{N1}\lt(\lt.\begin{array}{ll}a&b\\d&c\end{array}\rt|u\rt)$ was
given in \cite{Jim88}. So, we complete the proof. ~~~~~~~$\square$

\subsection{The ``roots of unity" case}
From now on, we come back to the case that the crossing parameter
takes value of ``roots of unity" $w=\frac{n}{N}$. It was shown in
\cite{Bel02} that for the eight-vertex (or $\Zb_2$) case  the
subspace $Ker(S_N)$ is larger than $W_0$ by an additional
$2$-dimensional subspace in ``roots of unity" case. We shall show
that $Ker(S_N)$ is also larger than $W_0$ by an additional
$n$-dimensional subspace for the generic $\Zb_n$ case.

Let us introduce $n$ vectors $\Psi_a^i(u)\in
V_1\otimes\cdots\otimes V_N$, $i=1,\cdots,n,$  \bea
\Psi_a^i(u)=\Phi_{N;a,a+Ne_i}(u),~~{\rm for~a~generic~
vector~}a\in \Cb^n.\label{Bas} \eea We denote the $n$-dimensional
subspace spanned by the vectors $\{\Psi_a^i(u)|i=1,\cdots,n\}$ by
$\overline{W}$ which is independent of the choice of $a$.

\begin{Proposition} \label{Ker}
When the crossing parameter $w=\frac{n}{N}$, $
Ker(S_N)=\overline{W}\oplus W_0$.
\end{Proposition}
{\it Proof\/}. Using the Yang-Baxter equation (\ref{YBE-V}), one
can permute the order of the factors of $S_N$ in such way \bea
S_N=\cdots R_{1N}((N-1)w)R_{1N-1}((N-2)w)\cdots R_{12}(w).\no\eea
Acting it on the vector $\phi_{N;a,a+Ne_i}^{i\cdots i}(u)$ defined
in (\ref{Ph-1}) (There exists a unique $N$-path for the $N$-{\it
admissible\/} pair $(a,a+Ne_i)$ such as
$(a,a+e_i,a+2e_i,\cdots,a+Ne_i)$) and using Theorem \ref{VF}, one
can derive \bea &&S_N\phi_{N;a,a+e_i}^{i\cdots i}(u)=\cdots
R_{1N}((N-1)w)R_{1N-1}((N-2)w)\cdots
R_{12}(w)\phi_{N;a,a+Ne_i}^{i\cdots i}(u)\no\\
&&~~~~=\cdots R_{1N}((N-1)w)\cdots
R_{12}(w)\phi_{a,a+e_i}(u+(N-1)w)\otimes\cdots\phi_{a+(N-1)e_i,a+Ne_i}(u)\no\\
&&~~~~=\cdots\times\frac{\s(Nw)\s((N-1)w)\cdots\s(w)}{(\s(w))^N}
\phi_{a+(N-1)e_{i},\,a+Ne_{i}}(u+(N-1)w)\no\\
&&~~~~~~~~\otimes\phi_{a,\,a+e_{i}}(u+(N-2)w)\otimes\cdots\otimes
\phi_{a+(N-2)e_{i},\,a+(N-1)e_{i}}(u).\eea The fact that
$\phi_{N;a,a+Ne_i}^{i\cdots i}(u)\in Ker(S_N)$ follows from the
identity $\s(Nw)=\s(n)=0$. Noting that \bea
\Psi_a^i(u)=\phi_{N;a,a+Ne_i}^{i\cdots i}(u)~~{\rm
mod}~W_0,\no\eea and the fact $W_0\subset Ker(S_N)$, one can show
$\overline{W}\subset Ker(S_N)$. Evidently $\overline{W}\cap
W_0=\{0\}$. It follows that $\overline{W}\oplus W_0\subset
Ker(S_N)$. Then it remains to prove that \bea {\rm dim}~
Ker(S_N)={\rm dim}~ W_0\,+\, {\rm dim}~
\overline{W}.\label{D-I}\eea However, the dimensions of these
subspaces do not depend on the modulus $\tau$ of the torus. Taking
the trigonometric limit $\tau\longrightarrow +\sqrt{-1}\infty$,
the equation (\ref{D-I}) has been already proved in \cite{Tar93}.
So we complete the proof. ~~~~~~$\square$

Proposition \ref{Invar} and \ref{Ker} show that $\overline{W}$ is
invariant under the fused  R-matrix ${\cal{R}}^{(s)}_{1\cdots
N}(u)$ defined in (\ref{F-R1}) and the L-operator
${\cal{L}}^{(s)}_{1\cdots N}(u)$ defined in (\ref{F-L1}). Let
$\hat{R}(u)\in {\rm End}(\overline{W}\otimes \Cb^n)$ and $\L(u)\in
{\rm End}(\overline{W})\otimes \A$ be the restrictions of the
fused operators ${\cal{R}}^{(s)}_{1\cdots N}(u)$ and
${\cal{L}}^{(s)}_{1\cdots N}(u)$ on $\overline{W}$, explicitly,
\bea &&{\cal{R}}^{(s)}_{1\cdots N}(u)\Psi_a^i(u)=
\hat{R}(u)\Psi_a^i(u)=\sum_{j=1}^n\hat{R}_j^i(u)\Psi_a^j(u),~~
\hat{R}^i_j(u)\in{\rm End}\,(\Cb^n), \label{F-R2}\\
&&{\cal{L}}^{(s)}_{1\cdots
N}(u)\Psi_a^i(u)=\L(u)\Psi_a^i(u)=\sum_{j=1}^n\L_j^i(u)\Psi_a^j(u),~~\L_i^j(u)\in
\A .\label{F-L2}\eea

\section{The Center }
\label{CE} \setcounter{equation}{0}

Let us compute some of the fused Boltzmann weights with a
specially chosen configuration $a,b,c,d\in \Cb^n$. By using
Proposition \ref{F-W-P} and the identity
$\s(Nw\d_{i\nu})=\s(n\d_{i\nu})=0$,  we can derive for the
crossing parameter $w=\frac{n}{N}$ \bea
&&W_{N1}\lt(\lt.\begin{array}{ll}a&a+e_{\nu}\\a+
Ne_i&a+Ne_i+e_{\mu}\end{array}\rt|u\rt)\no\\
&&~~~~= \lt(\prod_{j=1}^{N-1}\frac{\s(u+jw)}{\s(w)} \rt)
\frac{\s(u+Nw\d_{i\mu})\prod_{\r\neq
\mu}\s(a_{\r\mu}w+Nw\d_{i\r})}
{\prod_{\r}\s(a_{\r\mu}w+\d_{\r\mu}w)}\d_{\mu\nu},\no\eea in which
the quasi-periodicity of the $\s$-function \bea
\s(u+Nw)=\s(u+n)=(-1)^n\s(u),~~Nw=n,\eea is used. This leads to a
simple expression of the fused Boltzmann weight \bea
&&W_{N1}\lt(\lt.\begin{array}{ll}a&a+e_{\nu}\\a+Ne_i&a+Ne_i+e_{\mu}\end{array}\rt|u\rt)=
\lt(\prod_{j=1}^{N}\frac{\s(u+jw)}{\s(w)}
\rt)\d_{\mu\nu}.\label{Main-i}\eea In order to compute the
restricted R-matrix $\hat{R}(u)$ defined in (\ref{F-R2}), we
evaluate the action of the fused R-matrix
${\cal{R}}^{(s)}_{1\cdots N}$ defined in (\ref{F-R1}) on the
vector $\Psi_a^i$ \bea &&{\cal{R}}^{(s)}_{1\cdots
N;0}(u-v)\Psi_a^i(u)\otimes
\phi_{a+Ne_i,a+Ne_i+e_j}(v)\no\\
&&~~~~=\lt(\prod_{k=1}^{N}\frac{\s(u-v+kw)}{\s(w)}\rt)~
\Psi_{a+e_j}^i(u)\otimes \phi_{a,a+e_j}(v).\eea Using the shift
property of the fundamental intertwining vector
$\phi_{\l,\l+e_i}(u)$ (\ref{F-S-P}), one can write the fused
intertwining vector explicitly as \bea
&&\Psi_a^i(u)=P^{(+)}_N\lt\{\mbox{\vphantom{\Huge L}}
\phi_{a,a+e_i}(u+(N-1)w)\otimes
\phi_{a,a+e_i}(u+(N-1)w-nw)\rt.\no\\[5pt]
&&~~~~~~~~~~~~\otimes\cdots\otimes\lt.\phi_{a,a+e_i}(u+(N-1)w-(N-1)nw)
\mbox{\vphantom{\Huge L}}\rt\}.\label{F-P-2}\eea The periodicity
(\ref{G-P})  enables one to further derive the following relation
\bea &&\phi_{a,a+e_j}(u+Nw)=\phi_{a,a+e_j}(u+n)
=\phi_{a,a+e_j}(u).\label{Eq-1}\eea The above equation and
(\ref{F-P-2}) imply the following relations
 \bea
&&\Psi_{a}^i(u+nw)=\Psi_a^i(u), \label{F-P-3}
\\&&\Psi_{a+e_j}^i(u)=\Psi_a^i(u+w).
\label{F-P-4}\eea Then
\begin{Proposition} \label{Main-L}When the crossing parameter is in ``roots of
unity" case $w=\frac{n}{N}$, the vector $\Psi_{a}^i(u)$ satisfies
the following relation \bea
\Psi_{a+e_j}^i(u)=\Psi_a^i(u+w)=\Psi_a^i(u).\eea
\end{Proposition}
{\it Proof\/}. The first equality is just equation (\ref{F-P-4}).
We are to prove the second equality $\Psi_a^i(u+w)=\Psi_a^i(u)$.
The expression of $\Psi_a^i(u)$ given in (\ref{F-P-2}) and
equation (\ref{Eq-1}) imply that \bea
\Psi_a^i(u+Nw)=\Psi_a^i(u+n)=\Psi_a^i(u).\label{Eq-2}\eea Due to
the fact that $N$ and $n$ are positive coprime integers,  one can
choose an  integer $l$ such that \bea 1+ lN =0~{\rm mod}~ n.\eea
The above property and relations (\ref{F-P-3}) and (\ref{Eq-2})
allow one to derive that \bea \Psi_a^i(u+w)=\Psi_a^i(u+w+
lNw)=\Psi_a^i(u+(1+ lN)w)=\Psi_a^i(u).\eea Hence, we complete the
proof. ~~~~~~$\square$

\noindent Using Proposition \ref{Main-L}, the action of the fused
R-matrix ${\cal{R}}^{(s)}_{1\cdots N}$ on the vector $\Psi_a^i$
can be given explicitly as
\begin{Proposition}\label{Main-Prop}
\bea &&{\cal{R}}^{(s)}_{1\cdots N;0}(u-v)\Psi_a^i(u)\otimes
\phi_{a+Ne_i,a+Ne_i+e_j}(v)\no\\
&&~~~~=\lt(\prod_{k=1}^{N}\frac{\s(u-v+kw)}{\s(w)}\rt)~
\Psi_{a}^i(u)\otimes \phi_{a,a+e_j}(v).\eea
\end{Proposition}

Now, we calculate the exchange relation among the fused
L-operators $\{\L^{l}_{k}(u)|\,k,l=1,\cdots,n\}$ and the
generators of $\A$ $\{L^j_i(u)|\,i,j=1,\cdots,n\}$. Firstly, one
can derive the following relation  by the  ``RLL" relation
(\ref{RLL}) \bea R_{1\cdots N;0}(u-v)L_{1\cdots
N}(u)L_0(v)=L_0(v)L_{1\cdots N}(u)R_{1\cdots N;0}(u-v).\eea
Proposition \ref{Invar} allows one to restrict the above equation
on $W_0^{\perp}$, namely in terms of the fused R-matrix
(\ref{F-R1}) and L-operator (\ref{F-L1}) \bea
{\cal{R}}^{(s)}_{1\cdots N;0}(u-v){\cal{L}}^{(s)}_{1\cdots
N}(u)L_0(v)=L_0(v){\cal{L}}^{(s)}_{1\cdots
N}(u){\cal{R}}^{(s)}_{1\cdots N;0}(u-v).\label{Main-E-R}\eea
Proposition \ref{Ker} and the fact that $\overline{W}\subset
W_0^{\perp}$ allow one further to restrict the equation on
$\overline{W}$. Finally, we have our main result.
\begin{Theorem}\label{Main-T}When the crossing parameter is in ``roots of
unity" case $w=\frac{n}{N}$, the elements
$\{\L^i_j(u)|\,i,j=1,\cdots,n\}$ defined in (\ref{F-L2}) are the
central elements of $\A$, namely, \bea [\L^i_j(u),\,
L^k_l(v)]=0.\eea
\end{Theorem}
{\it Proof\/}.  For a generic $a\in \Cb^n$, we can introduce an
$n\times n$  matrix $A$ with matrix elements $A^i_j$: \bea
A^i_j=\phi_{a,a+e_j}^{(i)}(u), ~~i,j=1,\cdots,n.\no\eea One can
verify that $det(A)\neq 0$. It means that \bea {\rm
Span}~\{\phi_{a,a+e_j}(u)|\,j=1,\cdots,n\}= \Cb^n
.\label{Comp-1}\eea This fact allows one to define matrix elements
$\tilde{L}^j_i(u)\in \A$ from the L-operator $L(u)$ as follows
\bea
L(u)\phi_{a,a+e_i}(u)=\sum_{j=1}^n\tilde{L}^j_i(u)\phi_{a,a+e_j}(u).
\label{Def3}\eea The conditions $w=\frac{n}{N}$ and (\ref{G-P})
allow one to derive the following relation \bea
&&\phi_{a+Ne_i,a+Ne_i+e_j}(u)=\phi_{a,a+e_j}(u+Nw)
=\phi_{a,a+e_j}(u).\label{P-F-P}\eea Acting both sides of equation
(\ref{Main-E-R}) on the vector $\Psi_{a}^l(u)\otimes
\phi_{a,a+e_j}(v)$ and using the definitions (\ref{F-L2}) and
(\ref{Def3}), we have\bea &&{\rm LHS}=\sum_{k,i=1}^n
{\cal{R}}^{(s)}_{1\cdots N;0}(u-v)\Psi_{a}^k(u)\otimes
\phi_{a,a+e_i}(v)\{\L_k^l(u)\tilde{L}^i_j(v)\},\no\\
&&{\rm RHS}=L_0(v){\cal{L}}^{(s)}_{1\cdots N}(v)
{\cal{R}}^{(s)}_{1\cdots N;0}(u-v)\Psi_{a}^l(u)\otimes
\phi_{a,a+e_j}(v).\no \eea Proposition \ref{Main-Prop} and the
relation (\ref{P-F-P}) imply that we can further derive that \bea
&&{\rm
LHS}=\lt\{\prod_{k=1}^N\frac{\s(u-v+kw)}{\s(w)}\rt\}\sum_{k,i=1}^n
\Psi_{a}^k(u)\otimes
\phi_{a,a+e_i}(v)\{\L_k^l(u)\tilde{L}^i_j(v)\},\no\\
&&{\rm RHS}=
\lt\{\prod_{k=1}^N\frac{\s(u-v+kw)}{\s(w)}\rt\}\sum_{k,i=1}^n
\Psi_{a}^k(u)\otimes \phi_{a,a+e_i}(v)
\{\tilde{L}^i_j(v)\L_k^l(u)\} .\no \eea Hence
$[\L_k^l(u),\,\tilde{L}^i_j(v)]=0$. Equation (\ref{Def3}) and
non-degeneracy of the matrix $(A_j^i)$ enable us to derive that
$[\L_k^l(u),\,L^i_j(v)]=0$. Therefore, we complete the proof.
~~~~~~$\square$

\section{Conclusions}
\label{Con} \setcounter{equation}{0} In this paper we have
investigated the {\it extra\/} central elements (differing from
the so-called quantum determinant $det_q(L(u))$) of the algebra
$\A$ of monodromy matrices associated with the $\Zb_n$ R-matrix
when the crossing parameter takes special value $w=\frac{n}{N}$.
In the trigonometric limit, this case corresponds to the quantum
group $U_q(gl_n)$ at roots of unity. Our results show that in even
elliptic case the center of $\A$ is also extended and is generated
by $n^2$ elements $\{\L^i_j(u)\}$ in addition to  $det_q(L(u))$.
For the special case of $n=2$, our result recovers that of
\cite{Bel02}.

\section*{Acknowledgements}
W.\,-L. Yang is supported from  the Japan Society for the
Promotion of Science and Australian Research Council. A. Belavin is
partially supported by grant
RFBR-04-02-16027 and NATO grant PST.CLG.979008. He also would like
to thank Yukawa Institute for Theoretical Physics, Kyoto
University and Professor R. Sasaki for kind hospitality during his
visit to Kyoto.

\section*{Appendix: Proof of Proposition \ref{Prop-P}}
\setcounter{equation}{0}
\renewcommand{\theequation}{A.\arabic{equation}}
The property (\ref{F-S-P}) allows one to derive the following
relation \bea
&&\phi_{\l+e_j,\l+e_i+e_{j}}(u)=\phi_{\l,\l+e_{i}}(u+w),~~{\rm
if}~ i\neq j.\label{A-1}\eea For a vector $B\in
V_1\otimes\cdots\otimes V_m$, the corresponding symmetrized vector
$B^{(s)}$ is given as \bea
B^{(s)}=P^{(+)}_m(B)=\frac{1}{m!}\{\sum_{P\in\P_m}P(B)\}.\no\eea
The terms obtained by non-trivial permutations of $B$ are called
{\it descendants\/}. It is easy to show that the symmetrized
vector satisfies \bea B^{(s)} =P^{(+)}_m(PB),~~\forall P\in
\P_m.\label{A-2}\eea Hereafter we shall always keep equations
(\ref{A-1}) and (\ref{A-2}) in mind, because they will play an
important role to proof the proposition.

Let us denote the vector
$\phi_{m;a,a+\sum_{l=1}^me_{i_l}}^{i_1\cdots i_m}$ by $i_1\cdots
i_m$, and the set of
$\{\phi_{m;a,a+\sum_{l=1}^me_{i_l}}^{i_{p_1}\cdots
i_{p_m}}|\,P\in\P_m\}$ by $\overbrace{i_1\cdots i_m}$. We shall
prove the proposition by induction.

\begin{itemize}
\item  Using the relation (\ref{A-1}), we can prove that the
equation (\ref{Main-E}) holds for the case of $N=2$.
\item Suppose that (\ref{Main-E}) holds for the case
of $2<N$. We are to prove it is satisfied for $N+1$.

\begin{enumerate}
\item In the case of $i_{N+1}\not\in \{i_1,\cdots,i_N\}$, \bea
&&\Phi_{N+1;a,a+\sum_{l=1}^{N+1}e_{i_l}}^{i_1\cdots
i_{N+1}}\propto
P^{(+)}_{N+1}\lt\{\begin{array}{cc}\overbrace{i_1\cdots
i_N}&~~i_{N+1}\\\vdots&\end{array} \rt\},\label{A-3}\eea where
$\vdots$ stands for the {\it descendants\/}. Using the induction
hypothesis and (\ref{A-1}) we can derive that (\ref{A-3}) is equal
to \bea
&&P^{(+)}_{N+1}\lt\{\hspace{-1mm}\begin{array}{cc}i_1\cdots
i_{N}&i_{N+1}\\\vdots&\end{array}\hspace{-1mm}+\hspace{-1mm}
\begin{array}{cc}i_1\cdots
i_{N+1}&i_{N}\\\vdots&\end{array}\hspace{-1mm}+\hspace{-1mm}
\begin{array}{cc}i_1\cdots
i_{N+1}&i_{N-1}\\\vdots&\end{array}\hspace{-1mm}+\hspace{-1mm}
\cdots \rt\} \no\\[5pt]
&&~~~~~~=P^{(+)}_{N+1}\lt\{\overbrace{i_1\cdots
i_{N+1}}\rt\}.\no\eea So, (\ref{Main-E}) holds for  $N+1$.
\item If $i_{N+1}\in\{i_1,\cdots,i_N\}$, without loss of
generality, let us  suppose that $i_{N+1}=i_N$.  Let
$[i_N]^{\perp}$ denote the set $\{i_l\neq i_N|\,l=1,\cdots N-1\}$.
Let $M$ be the number of elements in the set. Let
$[i_N]^{\perp}_l$, $l=1,\cdots,M$, denote the element of the set.
Similarly to the first case, we have
 \bea
&&\Phi_{N+1;a,a+\sum_{l=1}^{N+1}e_{i_l}}^{i_1\cdots
i_{N}\,i_N}\propto
P^{(+)}_{N+1}\lt\{\begin{array}{cc}\overbrace{i_1\cdots
i_N}&i_{N}\\\vdots&\end{array} \rt\}\no\\[5pt]
&&~~=P^{(+)}_{N+1}\lt\{\begin{array}{cc}i_1\cdots
i_{N}&i_{N}\\\vdots&\end{array}+\begin{array}{cc}i_1\cdots
i_{N}&[i_{N}]^{\perp}_1\\\vdots&\end{array}+\cdots+
\begin{array}{cc}i_1\cdots
i_{N}&[i_{N}]^{\perp}_{M}\\\vdots&\end{array}\rt\} \no\\[5pt]
&&~~=P^{(+)}_{N+1}\lt\{\overbrace{i_1\cdots i_{N}~i_{N}}\rt\}.\no
\eea Then (\ref{Main-E}) holds for $N+1$.
\end{enumerate}
\end{itemize}
Therefore (\ref{Main-E}) holds for any positive integer $N\geq 2$.
Finally we complete the proof of Proposition \ref{Prop-P}.


\end{document}